\documentclass[aps,prl,twocolumn,showpacs,amsmath,floatfix]{revtex4}

\usepackage{graphicx}
\usepackage{dcolumn}

\def\ee{\boldsymbol{\cal{E}}}

\begin{document}


\title{
Wannier-based definition of layer polarizations in perovskite superlattices
}

\author{
Xifan Wu,
Oswaldo Di\'eguez,
Karin M. Rabe,
and David Vanderbilt
}

\affiliation{
Department of Physics and Astronomy, Rutgers University,
Piscataway, New Jersey 08854-8019, USA
}

\date{\today}

\begin{abstract}

In insulators, the method of Marzari and Vanderbilt [Phys.~Rev.~B
{\bf 56}, 12847 (1997)] can be used to generate maximally localized
Wannier functions whose centers are related to the electronic
polarization.  In the case of layered insulators, this approach can
be adapted to provide a natural definition of the local
polarization associated with each layer, based on the locations of
the nuclear charges and one-dimensional Wannier centers comprising
each layer.  Here, we use this approach to compute and analyze layer 
polarizations of ferroelectric perovskite superlattices, including 
changes in layer polarizations induced by sublattice displacements 
(i.e., layer-decomposed Born effective charges) and local symmetry 
breaking at the interfaces.  The method provides a powerful tool 
for analyzing the polarization-related properties of complex layered 
oxide systems.

\end{abstract}


\maketitle


\marginparwidth 2.7in

\marginparsep 0.5in

\def\dvm#1{\marginpar{\small DV: #1}}
\def\xwm#1{\marginpar{\small XW: #1}}
\def\kr#1{\marginpar{\small KR: #1}}
\def\od#1{\marginpar{\small OD: #1}}


Multicomponent superlattices based on the ABO$_3$ perovskite
structure have
received much attention recently due to the exciting properties they
possess as multifunctional materials (see Ref.~\onlinecite{Ghosez} and
references therein).
Experimental studies using modern layer-by-layer epitaxial growth techniques
have gone hand in hand with accurate first-principles calculations that
have helped to interpret experimental results and to guide the search
for superlattice compounds with tailored properties.
For example, a compositional perturbation that breaks inversion symmetry
was predicted \cite{Sai} to allow tuning of the dielectric and
piezoelectric
response, as confirmed later when such superlattices could be grown
experimentally \cite{Warusawithana, Lee}. In addition,
strained-layer superlattices can show a substantial enhancement of
spontaneous polarization; such effects have been observed
\cite{Lee} and analyzed using first-principles calculations
\cite{Jeff,SergeI}.

One issue that has received much attention theoretically
is how to quantify the concept of {\it local} polarization.
This can be very useful in isolating the contributions of
constituent layers to dielectric and piezoelectric
properties \cite{Sai,Warusawithana}, separating the effects of
factors such as epitaxial strain and applied electric fields
\cite{SergeI,Johnston,Jeff,Sepliarsky}, and understanding
the enhancement or suppression of spontaneous polarization
\cite{Jeff,SergeI}.  A local description is also
essential for characterizing and understanding
interface contributions to such properties. Among
previously proposed local approaches \cite{Sepliarsky,Meyer}, that
of Meyer and Vanderbilt \cite{Meyer} has been one of the most
commonly used \cite{SergeI,Johnston,Jeff}.  Based on a linear
approximation involving effective charges and small ionic
distortions from a higher-symmetry nonpolar reference structure,
this simple model captures the essential physics and provides a
semi-quantitative description useful for understanding many aspects
of the behavior of multicomponent superlattices.  However, as we
shall discuss, it is neither exact nor unique.

A first-principles method for identifying local dipoles and
computing their dipole moments in an extended system has been
proposed in Refs.~\cite{Giustino03,Giustino05}, based on expressing
the electric polarization in terms of the centers of charge of
Wannier functions (WF) \cite{King-Smith,Marzari} that are maximally
localized along the direction of interest \cite{SPR}.  The
method was successfully applied to analyze the permittivity
of ultra-thin Si-SiO$_2$ heterostructures
\cite{Giustino03,Giustino05}.

Here, we further develop a closely related method and apply them to
ferroelectric perovskite 
superlattices, where the polarization, particularly that along the growth
direction $z$, is of central physical importance. We introduce a WF-based 
expression for the ``layer polarization" along $z$ 
associated with each charge-neutral AO
or BO$_2$ layer in an (001) superlattice built from II-VI ABO$_3$
perovskites such as BaTiO$_3$, SrTiO$_3$, and PbTiO$_3$.
Unlike the approach of Ref.~\onlinecite{Meyer}, the
present one is exact (i.e., the sum of LPs relates exactly
to the total supercell polarization) and is entirely free of
arbitrary choices in its implementation.  We will present examples
showing how this approach naturally provides an insightful local
description of the polarization behavior of perovskite
superlattices, both at zero electric field and under nonzero
electrical bias, and in particular yields valuable information
about the highly localized atomic and electronic rearrangements at
the interfaces.

The modern theory of polarization \cite{King-Smith} is routinely used
to compute the polarization of a crystal as a sum of ionic and electronic
(Berry phase) contributions.  In the Wannier representation, this
takes the form
\begin{equation}
{\bf P}=\frac{1}{V} \sum_\tau Q_\tau\,{\bf R}_\tau -\frac{2e}{V}
\sum_m \bar{\bf r}_m
\end{equation}
where $\tau$ and $m$ run over ion cores (of charge $Q_\tau$ located
at ${\bf R}_\tau$) and Wannier centers (of charge $-2e$ located
at $\bar{\bf r}_m$), respectively, in the unit cell of volume $V$.
In the case of a II-VI perovskite superlattice, one may hope to
decompose the system into neutral ``layers'' (that is,
AO or BO$_2$ subunits) and define a ``layer polarization'' (LP)
\begin{equation}
p_j=\frac{1}{S} \sum_{\tau\in j} Q_\tau\,R_{\tau z} -\frac{2e}{S}
\sum_{m\in j} \bar{z}_m
\label{eq:layers}
\end{equation}
in which the sums are restricted to entities belonging to layer
$j$.  Here $S$ is the basal cell area and we are now focusing only
on $z$ components.  The LP $p_j$ thus defined has
units of dipole moment per unit area, and the total polarization, with units
of dipole moment per volume, is
exactly related to the sum of LPs via
$P_z={c}^{-1}\sum_j p_j$ where $c=V/S$ is the supercell lattice
constant along $z$.
For such a decomposition to be meaningful, we need (i) to resolve the
arbitrariness associated with the positions of the Wannier centers,
and (ii) to be satisfied that the Wannier centers can be assigned
to layers without ambiguity.  We shall show below by example that (ii)
is satisfied for the systems of interest, and thus we next turn
our attention to issue (i).

As is well known, the Wannier centers $\bar{\bf r}_m=\langle W_m \vert
{\bf r} \vert W_m \rangle$ are not unique because the electronic
structure is invariant to unitary rotations among the WFs (corresponding,
e.g., to different choices of phases of the Bloch functions in k-space).
Marzari and Vanderbilt \cite{Marzari} introduced a method for obtaining
a unique set of WFs by choosing the ones that minimize the sum of
second-moment spreads (spatial variances) of the WFs.  In a
three-dimensional system, this involves finding a best possible
compromise between minimal spread in $x$, $y$, and $z$ directions,
and an iterative procedure is needed to find this compromise
solution.

Here, we are interested only in polarizations along $z$, and
can limit ourselves to minimizing the spread only in that direction
\cite{Giustino03,Giustino05}.
Moreover, we can use a hybrid representation of the electronic ground state
that is Bloch-like in $x$ and $y$ and Wannier-like only along $z$
\cite{SPR}.
We start from a conventional band-structure calculation
carried out on a mesh of reciprocal points ${\bf k}=(k_x,k_y,k_z)$
and adopt the relabeling ${\bf g}=(k_x,k_y)$ and $k=k_z$.
That is, each 2D vector $\bf g$ labels a string of $J$ k-points
running along the $z$ direction with separation $b=2\pi/Jc$.  Our
task is then to transform the Bloch functions
$\vert\psi_{{\bf g},nk}\rangle$ ($n=1, \ldots, N$) into hybrid WFs
\cite{SPR}
$\vert W_{{\bf g},m}\rangle$ ($m=1, \ldots, N$) via a 1D Wannier transform,
where $N$ is the number of occupied bands.  We will then let $\bar{z}_m$
in Eq.~(\ref{eq:layers}) be the average of $\bar{z}_m({\bf g})=
\langle W_{{\bf g},m}\vert z\vert W_{{\bf g},m}\rangle$ over the 2D
mesh of ${\bf g}$ points.  Since the 1D Wannier transform is done
independently at each ${\bf g}$, we drop the ${\bf g}$ label in the
following paragraph.

The case of maximally localized WFs in 1D was treated explicitly
in Sec.~IV.C.1 of Ref.~\onlinecite{Marzari}.  There, it was shown that the
WFs that minimize the spread functional are identical to the
eigenfunctions of the projected position operator $PzP$, where
$P=\sum_{nk} \vert \psi_{nk} \rangle \langle \psi_{nk} \vert$
is the band projection operator.  It was also shown how they could be
obtained from a parallel-transport based construction using the
singular value decomposition (SVD) of the overlap matrices between
neighboring k-points, $M_{mn}^{(k)}=\langle u_{mk} \vert u_{n,k+b} \rangle$,
where $u_{mk}$ is the periodic part of the Bloch function $\psi_{mk}$.
The SVD is
$M= V \Sigma W^{\dag}$ where $V$ and $W$ are unitary and $\Sigma$ is
positive real diagonal.  For small $b$, $\Sigma$ approaches the unit
matrix, and omitting $\Sigma$ to write $\widetilde{M}=UV^\dagger$
can be regarded as a way of constructing a purified version of
$M$ that is exactly unitary.  Then
$\Lambda=\prod_{k=1}^J \widetilde{M}^{(k)}$ defines a global
unitary matrix describing the parallel transport of the states
on the k-point string, and its unimodular eigenvalues $\lambda_m$
define the Wannier centers via $\bar{z}_m=(-c/2\pi)
\,{\rm Im}\ln\lambda_m$.  Note that no iterative procedure is required;
these 1D Wannier locations can be obtained by a straightforward
small-matrix diagonalization.  A procedure that is similar in spirit,
but slightly different in detail, has recently been proposed
elsewhere \cite{Bhattacharjee}.

\begin{figure}
\includegraphics[width=2.3in]{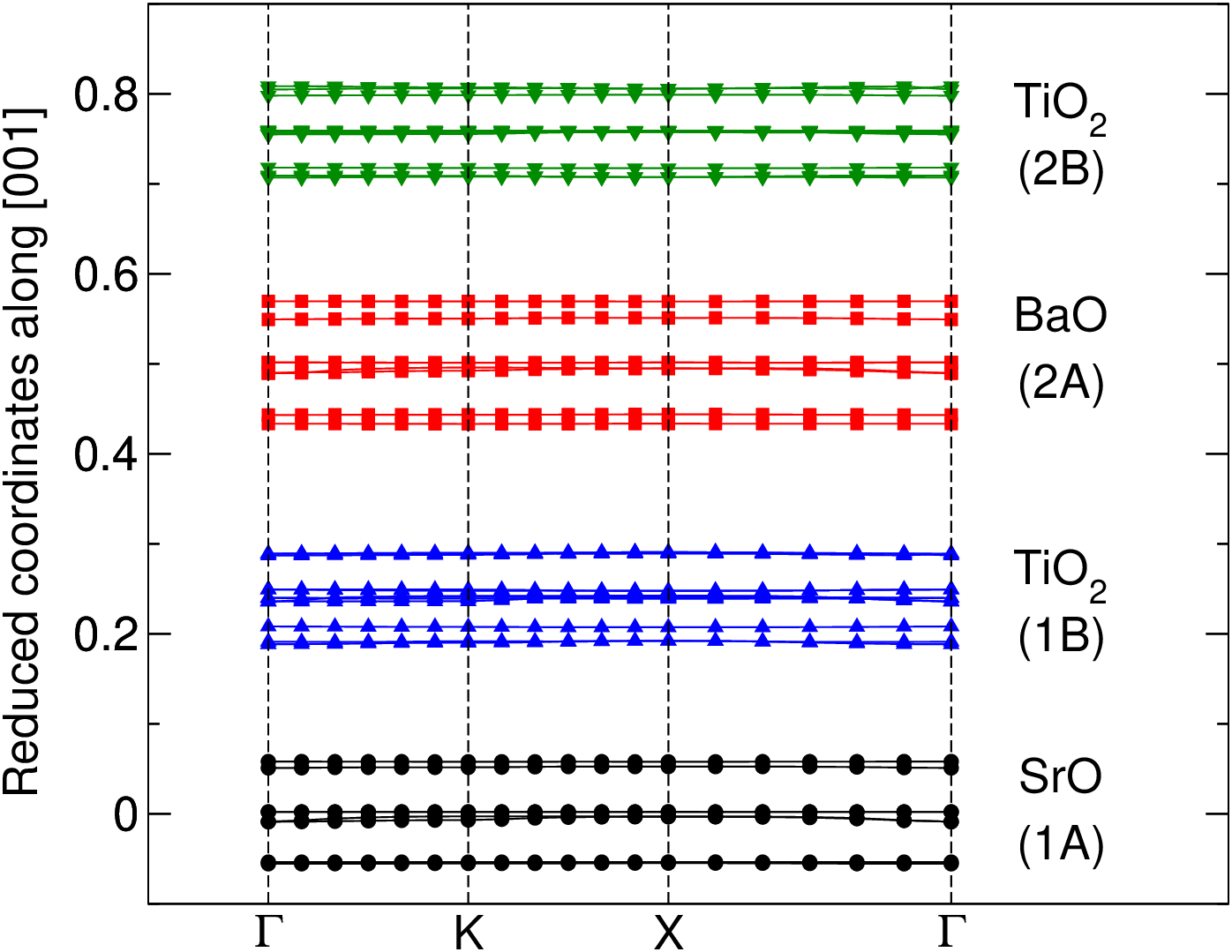}
\caption{\label{fig1} (Color online.)
Dispersion of WF center positions
for a 1BT/1ST superlattice as a function of ${\bf g}=(k_x,k_y)$.}
\end{figure}

In Fig.~(\ref{fig1}), we present the
resulting values of $\bar{z}_m({\bf g})$ for an {\it ab-initio}
calculation on a 10-atom tetragonal supercell composed of alternating
SrTiO$_3$ (ST) and BaTiO$_3$ (BT) units, which we refer to as a 1ST/1BT
superlattice. 
We did all calculations using the ABINIT code
\cite{ABINIT}, which implements density-functional theory within the
local-density approximation (LDA) \cite{LDA}.  We adopted Teter
norm-conserving pseudopotentials \cite{Teter} for which the valence
states are ($5s\,5p\,6s$) for Ba, ($4s\,4p\,5s$) for Sr, ($3s\,3p\,3d\,4s$)
for Ti, and ($2s\,2p$) for O.  We used a plane-wave energy cutoff of
45 Ha, a $6 \times 6 \times 3$ Monkhorst-Pack self-consistency mesh,
and a $12 \times 12 \times 3$ bandstructure mesh.  We assumed perfect
epitaxial growth of the superlattices on a cubic ST substrate having
a theoretical equilibrium lattice constant of 7.265 bohr and
tetragonal P4mm symmetry.

The key feature visible in the $\bar{z}$($\bf g$) dispersion relation
in Fig.~\ref{fig1} is that the WF centers
separate quite naturally into distinct layers as anticipated.
The 1D Wannier positions $\bar{z}$ are almost independent
of ${\bf g}=(k_x,k_y)$, and there are robust gaps between layers.
Moreover, we find eight Wannier centers in each BaO or SrO layer and
12 in each TiO$_2$ layer (four for each cation semicore shell and
four for each oxygen $2s2p$ shell), so that the layers are neutral
as expected.  All this demonstrates that the proposed
Wannier-based approach does indeed lead to a natural and robust
decomposition into easily identified neutral layers.

\begin{figure}
\includegraphics[width=3.0in]{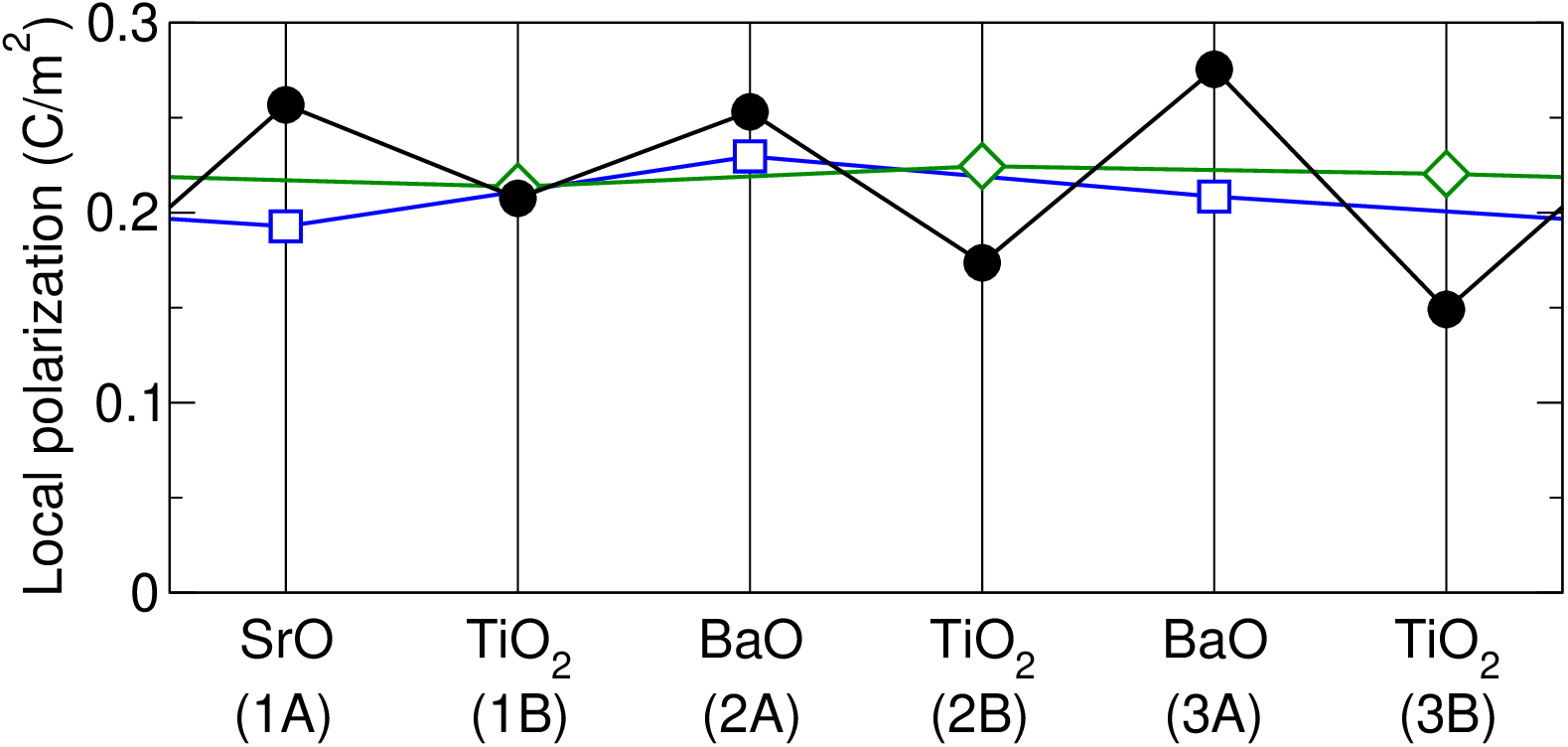}
\caption{\label{fig2} (Color online.)
Local polarization profile of 1ST/2BT supercell, from
effective charge approximation based on A centered (open square) or
B centered (open diamonds) analysis, and from layer polarization
analysis (filled circles). The overall supercell polarization
is 0.22\,C/m$^2$.}
\end{figure}

It is then straightforward to define the LP 
$p_j$ associated with each layer according to
Eq.~(\ref{eq:layers}).  For comparison with other definitions, we
also introduce the corresponding local polarization $P_j=p_j/c_j$ having
the correct units of polarization (dipole per unit volume), where
$c_j$ is chosen as half the distance between the two neighboring
cations.  In Fig.~\ref{fig2}, we show the local polarization
profile calculated in this way for the case of a 1ST/2BT
superlattice (15-atom supercell).  The details are the same as for
the calculation of the 1ST/1BT superlattice, except that we use a
$6 \times 6 \times 2$ k-point sampling.  We compare our results
with the ones obtained from the commonly applied
approximate scheme \cite{Meyer} in which the local polarization
is estimated by multiplying the Born effective charges
of the atoms in a unit cell layer by their displacements relative to a
reference structure.
The effective charges are obtained from
linear response calculations in the ferroelectric ground state.  By
its very nature, this approximate scheme \cite{Meyer} has only half
the spatial resolution of our new scheme, since it applies only to
entire ABO$_3$ cells.  (Note that it cannot easily be extended down
to the resolution of AO and BO$_2$ layers because the sum of $Z^\ast$
values in such a layer does {\it not} vanish, so that the definition
would depend on choice of reference structure in an unsatisfactory way.)

The results shown in Fig.~\ref{fig2} are consistent with the
findings of previous theoretical studies \cite{Jeff}
showing that the SrTiO$_3$ portion of the supercell becomes polarized
to almost the same degree as the BaTiO$_3$ portion.  However, the
improved resolution associated with the new approach is also clearly
evident in the figure.  For example, one can now see that the
polarization tends to be larger in the AO layers than in the TiO$_2$
layers (see next paragraph).
Moreover, our new approach is free of three
limitations of the approximate one \cite{SergeI,Jeff,Johnston,Meyer}.
First, we avoid the choice between an A- or B-centered analysis.
Second, we do not have the problem of choosing an arbitrary local
reference structure as the basis for the definition of the atomic
displacements.  Third, our LPs $p_j$ sum to give the exact total
polarization of the entire supercell, whereas the approximate ones
do not.

\begin{table}[b]
\caption{Layer decomposition of the [001] Born
effective charges in a 3BT supercell.  Total effective charges
are given in the last row.}
\begin{center}
\begin{tabular}{ldddd}
\hline \hline
& \multicolumn{1}{r}{Ti (1B)}
& \multicolumn{1}{r}{Ba (1A)}
& \multicolumn{1}{r}{O$_{\parallel}$ (1A)}
& \multicolumn{1}{r}{O$_{\perp}$ (1B)} \\
\hline
BaO (1A)      &  1.433  &  1.268  &  -2.448  &  -0.225  \\
TiO$_2$ (1B)  &  1.872  &  0.148  &  -0.231  &  -0.930  \\
BaO (2A)      &  1.262  &  0.434  &  -1.027  &  -0.191  \\
TiO$_2$ (2B)  &  0.619  &  0.296  &  -0.542  &  -0.216  \\
BaO (3A)      &  1.211  &  0.435  &  -1.046  &  -0.348  \\
TiO$_2$ (3B)  &  0.636  &  0.191  &  -0.264  &  -0.217  \\
\hline
$Z^\ast$     &  7.033  &  2.772  &  -5.557  &  -2.127  \\
\hline \hline
\end{tabular}
\end{center}
\label{table:zstar}
\end{table}

The Born effective charges $Z^{\ast}$, defined as the first
derivatives of polarization with respect to atomic displacements,
describe the dynamics of the charge transfer induced by
such displacements.  We now illustrate how the
LP concept can be used to decompose the $Z^\ast$ for an atom in one
layer into contributions from neighboring layers \cite{explan-GaAs}.
This is demonstrated in Table \ref{table:zstar} for the case of
a supercell of tetragonal bulk BT that
has been tripled along [001] (3BT supercell).
Each of the four symmetry-inequivalent atoms (Ti and
O$_{\parallel}$ in a TiO$_2$ layer and Ba and O$_{\perp}$ in a BaO
layer) was displaced along $z$ in turn, and the changes in all six 
LPs in the supercell were computed.  For
Ba, O$_{\perp}$ and O$_{\parallel}$, the induced polarizations are
dominated by contributions from the same atomic layer, at the level
of around 45\%.  In contrast, for the Ti atom, the contributions
from the first neighboring layers are almost as large as from the
layer itself.
This is consistent with the well-known role of the Ti($3d$)-O($2p$)
hybridization in giving rise to the anomalous Born effective
charges in these perovskites \cite{BTO}, and the fact that the
WFs that embody this hybridization reside on the O atoms.
Thus, a motion of the Ti atoms along [001] modulates this
hybridization and shifts the centers of the WFs residing on the
neighboring BaO layers.  This effect also helps explain why
the LPs for AO layers are larger than for
TiO$_2$ layers in Fig.~\ref{fig2}.

\begin{figure}
\includegraphics[width=3.2in]{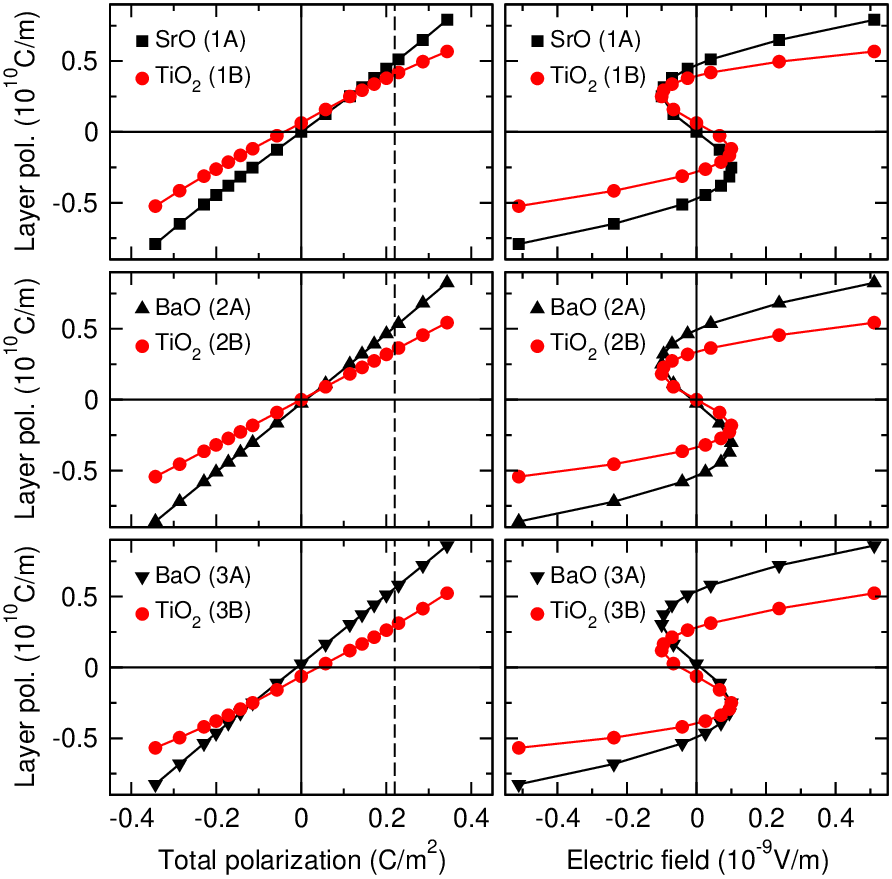}
\caption{\label{fig3} (Color online.) Layer polarization as a
function of (a) total polarization of the supercell, and (b) macroscopic
electric field in the supercell, for six consecutive layers in the
1ST/2BT supercell.  Labeling of layers follows Fig.~1.}
\end{figure}

We further illustrate the utility of the LP analysis by considering
its behavior in a macroscopic electric field $\ee$ \cite{Ivo} applied
along [001].  We used
a constrained-polarization mapping technique \cite{Oswaldo}
generalized to include volume relaxation \cite{Interface}
to find the minimum-energy configuration for each given
polarization.  The resulting LPs $p_j$ vs.\ total
polarization are shown for the 1ST/2BT supercell in the left-hand
panels of Fig.~\ref{fig3}.  We see that each LP is roughly
linear in the total polarization for all six layers, but with
nonlinearities appearing at large values of polarization.  For each
data point, we extracted the corresponding
macroscopic electric field value, and plotted the LPs against this
field in the right-hand panels of Fig.~\ref{fig3}.
The results show a strongly nonlinear
dependence, typical of that found for total polarization as a
function of electric field in ferroelectric materials.

The most striking features seen in Fig.~\ref{fig3} are for the two
interface layers (TiO$_2$ layers 1B and 3B). They show local breaking of
inversion symmetry; that is, the LP as a function of
macroscopic field does not pass through the origin, at which for
the system as a whole the symmetry of the $P4/mmm$ space group
requires $\text{\bf P} = 0$ and $\ee = 0$.  One of these interface
TiO$_2$ layers has a nearest-neighbor SrO layer above and a BaO layer
below, and the other vice versa, the two interface layers being
related by a mirror symmetry.  The LP approach give us much more
precise information about the response of these interfaces to applied
fields than could be obtained from an analysis of either the
total polarization, or of the local polarizations as previously defined
(involving a smearing over three sequential atomic layers).  We expect
that this method of analysis of interface layers will be invaluable
for identifying the interface contributions to the properties of
superlattices with more than two components, particularly those
with globally broken inversion symmetry \cite{Sai,Interface}.
Similar considerations apply to BaO layers 2A and 3A, which also see
an environment of broken inversion symmetry.  For these, however,
the symmetry breaking enters only at the level of second-neighbor
layers, so the effects are smaller in magnitude.

In summary, we have introduced a definition of the layer
polarization (LP) in a multicomponent perovskite superlattice
that is exactly related to the polarization of the full system and 
does not require choosing an arbitrary reference configuration.   
For each atomic layer, the LP is uniquely determined by the spatial
locations of ionic and WF centers, and it can easily be computed in
any first-principles code as a post-processing step after standard
electronic structure calculations.  Although this polarization is
not directly measurable experimentally, we show examples in which
the LP precisely quantifies polar distortions
throughout the superlattice, the high resolution of the definition
being particularly relevant for inspecting the behavior of interface
layers.  Immediate applications include modeling of
interface effects on total polarization of multicomponent
superlattices \cite{Interface}, systematic studies of self-poling
effects in superlattices \cite{Interface}, and studies of the coupling of
phonons to the interfaces. For superlattices containing magnetic
constituents, the spin degeneracy assumption can be relaxed, so that
the WF centers will have additional local spin ordering
information.


\acknowledgments

This work was supported by ONR Grant N00014-05-1-0054
and by the Center for Piezoelectrics by Design under ONR
Grant N00014-01-1-0365.




\begin{thebibliography}{0}

\bibitem{Ghosez}
Ph.~Ghosez and J.~Junquera,
in {\em Handbook of Theoretical and Computational Nanotechnology},
edited by M.~Rieth and W.~Schommers,
American Scientific (2006).

\bibitem{Sai}
N. Sai, B. Meyer, and D. Vanderbilt, Phys. Rev. Lett. {\bf 84}, 5636 (2000).

\bibitem{Warusawithana}
M. P. Warusawithana, E. V. Colla, J. N. Eckstein, and M. B. Weissman,
Phys. Rev. Lett. {\bf 90}, 036802 (2003).

\bibitem{Lee}
H. N. Lee, H. M. Christen, M. F. Chisholm, C. M. Rouleau, and D. H. Lowndes,
Nature (London) {\bf 433}, 395 (2005).

\bibitem{Jeff}
J. B. Neaton and K. M. Rabe, Appl. Phys. Lett. {\bf 82}, 1586 (2003).

\bibitem{SergeI}
S. M. Nakhmanson, K. M. Rabe, and D. Vanderbilt, Appl. Phys. Lett. {\bf 87}, 102906 (2005).


\bibitem{Johnston}
K. Johnston, X. Huang, J. B. Neaton, and K. M. Rabe, Phys. Rev. B
{\bf 71}, R100103(2005);





\bibitem{Sepliarsky}
M. Sepliarsky, M. G. Stachiotti, and R. L. Migoni,
Phys. Rev. Lett. {\bf 96}, 137603 (2006).



\bibitem{Meyer}
B. Meyer and D. Vanderbilt, Phys. Rev. B {\bf 65}, 104111 (2002).

\bibitem{Giustino03}
F. Giustino, P. Umari, and A. Pasquarello,
Phys. Rev. Lett. {\bf 91}, 267601 (2003).

\bibitem{Giustino05}
F. Giustino and A. Pasquarello,
Phys. Rev. B {\bf 71}, 144104 (2005).

\bibitem{King-Smith}
R. D. King-Smith and D. Vanderbilt, Phys. Rev. B {\bf 47}, R1651 (1993);
R. Resta, Rev. Mod. Phys. {\bf 66}, 899 (1994).

\bibitem{Marzari}
N. Marzari and D. Vanderbilt, Phys. Rev. B {\bf 56}, 12847 (1997).

\bibitem{SPR}
C. Sgiarovello, M. Peressi, and R. Resta,
Phys. Rev. B {\bf 64}, 115202 (2001).

\bibitem{Bhattacharjee}
J. Bhattacharjee and U.V. Waghmare, Phys. Rev. {\bf 71}, 045106 (2005).

\bibitem{ABINIT}
X. Gonze
{\it et al.},
Comput. Mater. Sci. {\bf 25}, 478 (2002).

\bibitem{LDA}
P. Hohenberg and W. Kohn, Phys. Rev. {\bf 136},
B864 (1964); W. Kohn and L. J. Sham, {\it ibid.} {\bf 140}, A1133 (1965).

\bibitem{Teter}
M. Teter, Phys. Rev. B {\bf 48}, 5031 (1993).

\bibitem{explan-GaAs} See also Sec.~VII of Ref.~\onlinecite{Marzari}
for a discussion of atom-by-atom decomposition of the $Z^*$ in GaAs.

\bibitem{BTO}
R. E. Cohen and H. Krakauer, Phys. Rev. B {\bf 42}, 6416 (1990);
Ph. Ghosez, J.-P. Michenaud, and X. Gonze, {\it ibid} {\bf 58}, 6224 (1998).

\bibitem{Ivo}
I. Souza, J. \'{I}\~{n}iguez, and D. Vanderbilt, Phys. Rev. Lett.
{\bf 89}, 117602 (2002).

\bibitem{Oswaldo}
O. Di\'eguez and D. Vanderbilt, Phys. Rev. Lett. {\bf 96}, 056401 (2006).

\bibitem{Interface}
X. Wu, O. Di\'eguez, K. M. Rabe, and D. Vanderbilt, 
in preparation.

\end{thebibliography}
\end{document}